\documentclass[12pt]{iopart}
\usepackage{epsfig}
\usepackage{amssymb}
\def\al{{\it et al}}

\begin{document}
\title{Sharp crossover and anomalously large correlation length in
driven systems}

\author{Y. Kafri, E. Levine, D. Mukamel and J. T\"or\"ok\footnote{Present address: Department of Theoretical
Physics, Budapest University of Technology, H-1111 Budapest,
Budafoki ut 8.} }
\address{Department of Physics of Complex Systems, Weizmann
Institute of Science, Rehovot, Israel 76100.}
\begin{abstract}
Models of one-dimensional driven diffusive systems sometimes
exhibit an abrupt increase of the correlation length to an
anomalously large but finite value as the parameters of the model
are varied. This behavior may be misinterpreted as a genuine phase
transition. A simple mechanism for this sharp increase is
presented. The mechanism is introduced within the framework of a
recently suggested correspondence between driven diffusive systems
and zero-range processes. It is shown that when the dynamics of
the model is such that small domains are suppressed in the
steady-state distribution, anomalously large correlation length
may build up. The mechanism is examined in detail in two models.
\end{abstract}

\pacs{05.40.-a, 02.50.Ey, 64.75.+g}

\maketitle

Driven diffusive systems have been studied extensively in recent
years~\cite{Zia}. Particular attention has been given to
one-dimensional models, which have been shown to behave very
differently from systems in thermal
equilibrium~\cite{Mukamel00,Schutz01}. For example, in contrast to
equilibrium systems with short range interactions, non-equilibrium
systems, characterized by dynamics which does not obey detailed
balance, may exhibit spontaneous symmetry breaking and phase
separation even when the dynamics is local.

It has recently been noted that numerical simulations may be
rather misleading in attempting to determine whether or not a
particular model exhibits phase separation. For example, direct
numerical simulations of a single-lane three-species model
introduced by Arndt \al~\cite{Arndt} (AHR) strongly indicate that
two phase transitions take place in the model. The first
transition is from a homogeneous state to a mixed state, in which
the system separates into two phases: one with a high density and
the other with a low density. In the second transition, which
takes place from the latter state, the system further phase
separates into three pure phases. Analytical studies of the model
\cite{Sasamoto} have shown that the mixed state suggested by the
numerical studies is in fact homogenous, and the model exhibits
only one transition, from the homogeneous to the three-phase
state. At the apparent transition to what was misinterpreted as a
mixed state the model exhibits a sharp increase of its correlation
length $\xi$ to an anomalously large but finite value. The change
in $\xi$ is very sharp and it occurs over a narrow range of the
parameters defining the model. Such a behavior may be easily
misinterpreted in numerical studies of systems smaller than $\xi$
as a genuine phase transition.

Recently a simple criterion for the occurrence of phase separation
in one-dimensional systems has been proposed \cite{Kafri}. This
criterion does not rely on direct numerical simulations, and it
enables one to determine whether or not a particular model
exhibits phase separation. It has been applied to a two-lane
driven model, introduced by Korniss \al~\cite{Korniss}. Direct
numerical simulations of this model have indicated that it
exhibits phase separation in a region of its parameter space.
Moreover, numerical simulations, to be discussed below, show what
seems like a sharp transition between a homogeneous and a phase
separated states as the parameters defining the model are varied.
On the other hand, the criterion conjectured in \cite{Kafri}
indicates that in fact this model, too, does not exhibit phase
separation. Thus the apparent phase separation observed in the
numerical studies may again be due to a sharp crossover which
results in a large but finite correlation length. The criterion
\cite{Kafri} addresses the issue of the existence of phase
separation however it does not provide an understanding as to why
the correlation length becomes anomalously large in certain cases,
and why the increase in $\xi$ takes place over such a narrow range
of model parameters.

In this Letter we suggest a simple mechanism which accounts for
the large correlation lengths and sharp crossover phenomena
observed in some driven systems. We show that when the dynamics is
such that small domains of the high density phase are suppressed
in the steady state distribution, the resulting correlation length
becomes anomalously large, leading to an apparent phase separation
in numerical studies of finite systems. The mechanism is discussed
within the framework put forward in \cite{Kafri}. In this
framework a correspondence is made between one-dimensional driven
systems and zero-range processes. We use this correspondence to
analyze the crossover phenomena observed in the AHR and the
two-lane models, and discuss the underlying mechanism in detail.

We start by briefly reviewing the physical framework of the
criterion introduced in \cite{Kafri}. In this framework the
dynamics of the driven system is modelled by a zero-range process
(ZRP). Such a process \cite{ZRP,Evans} is defined on a
one-dimensional lattice of $M$ sites, or ``boxes'', with periodic
boundary conditions. Particles, or ``balls'', are distributed
among the boxes, with the box $i$ occupied by $n_i$ balls. At each
time step a box $i$ is chosen at random and a ball is removed from
it and transferred to one of its nearest neighbors with rate
$w_{n_i}$. The rate $w_{n_i}$ depends only on the occupation
number, $n_i$, in that box. The steady-state weights of the ZRP
are known to have the form \cite{ZRP,Evans}
\begin{equation}
\label{eq:wbar} W_{\mbox{\footnotesize ZRP}} \left( \lbrace n_i
\rbrace \right) = \prod_{i=1}^{M}{z^{n_i}{\cal F}_{n_i}}\;,
\end{equation}
where $z$ is the fugacity, ${\cal F}_k = \prod_{m=1}^{k}{1/w_{m}}$
for $k \geq 1$, and ${\cal F}_0 =1$. The correspondence to driven
systems is made by identifying occupied boxes with domains of the
high density phase. Specifically, a configuration of the driven
system can be described as a sequence of high-density domains
separated by low-density intervals. In the ZRP, each high density
domain is represented by a box, and the number of balls in that
box corresponds to the domain length. The evolution of the driven
model may be studied by considering the currents flowing in and
out of the high density domains. This dynamics may be represented
by a ZRP, with the rate $w_{n}$ taken as the current $J_n$ leaving
a domain of size $n$. Within this picture, the existence of phase
separation in the driven model corresponds to a macroscopic
occupation of one of the boxes in the ZRP.


In \cite{Kafri} it was conjectured, based on the correspondence to
a ZRP, that the existence of phase separation is related to the
asymptotic large $n$ behavior of the currents $J_n$ on blocks of
size $n$. For $J_n$ of the form
\begin{equation}
\label{eq:Jn} w_n = J_n =
J_\infty\left(1+\frac{b}{n}+{O}\left(1/n^2\right)\right)\;,
\end{equation}
with $J_\infty>0$, the existence of phase separation depends only
on the coefficient $b$. Phase separation takes place at high
densities only if $b>2$, and does not take place at any density if
$b<2$. Moreover, when $J_n$ is asymptotically decreasing to zero
at large $n$, phase separation takes place at any density.


We now utilize the correspondence between the ZRP and driven
systems to gain insight into the sharp crossover phenomena
discussed above in the AHR and two-lane models. We begin by
considering the AHR model. This is a three-state model on a ring.
Each site is either empty ($0$), or occupied by a positive ($+$)
or a negative ($-$) particle. The model evolves by a random
sequential dynamics in which a pair of nearest neighbor sites is
chosen at random and exchanged with the following rates:
\begin{equation}
\label{eq:rates} +\;0 \mathop{\rightarrow}\limits^{\alpha}  0\;+
\;\;\;;\;\;\; 0\,- \mathop{\rightarrow}\limits^{\alpha}  -\,0
\;\;\;;\;\;\; +\,- \mathop{\rightleftharpoons}\limits^{1}_{q} -\,+
\;.
\end{equation}
This dynamics conserves the densities of particles of each type.
The two particle densities are taken to be equal. Exact
calculations within the grand-canonical ensemble \cite{Sasamoto}
have shown that this model has two states: a fully ordered state
for $q>1$, in which the system strongly phase separates into three
phases ($+$, $-$ and $0$), and a disordered state for $q<1$ where
particles and vacancies are homogeneously distributed. In the
regime $q<1$ the current is an analytical function of $q$.
However, in the vicinity of a particular value $q_0<1$ (which
depends on $\alpha$ and the density $\rho$) a sharp crossover
takes place from a regime in which the correlation length $\xi$ is
relatively small ($q<q_0$) to a regime in which $\xi$ is
anomalously large, but finite ($q_0<q<1$). In this regime the
correlation length was found to reach values of the order
$10^{70}$ for some range of the parameters. This makes it clear
why numerical simulations of chains of length of the order of a
few thousands suggest a phase-separated state for $q_0<q<1$.

To make the correspondence of the AHR model to the ZRP,
high-density domains are identified as uninterrupted sequences of
positive and negative particles, bounded by vacancies. The number
of boxes $M$ in the ZRP is equal to the number of vacancies in the
AHR model, and the number of balls $N$ is equal to the number of
positive and negative particles. Hence the mean occupation of a
box in the ZRP, $\phi=N/M$, is related to the density $\rho$ in
the AHR by $\rho=\phi/(1+\phi)$. The dynamics within a domain is
given by the rates for the positive and negative particles
exchange in (\ref{eq:rates}). The flow of particles in and out of
a domain is controlled by the exchange rate $\alpha$ of particles
with vacancies.

The dynamics within each domain is thus given by the well studied
partially asymmetric exclusion process
(PASEP)~\cite{Sasamoto99,Blythe00} with particles injected and
ejected from the boundaries with rate $\alpha$. For this process
the asymptotic form of the current is known to take the form
(\ref{eq:Jn}) with $J_\infty = (1-q)/4$. The coefficient $b$ of
the leading order correction is $3/2$ for $q<1-2\alpha$, and $-1$
for $q>1-2\alpha$. Note that in the case $b=-1$ the current is an
increasing function of the domain size. Therefore small domains
are more stable than large ones, and large correlation lengths
cannot build up. We therefore concentrate only on the regime
$b=3/2$, where the sharp crossover has been observed.

It has been noted that the domain size distribution of the AHR
model is exactly given by that of the corresponding ZRP
\cite{Kafri}. Taking $w_n$ to order $1/n$ it is straightforward to
show that for large $n$ the domain size distribution is given by
\begin{equation}
\label{eq:dist} P(n) \sim
\frac{1}{n^b}\exp(-n/\xi)\;\;;\;\;\xi=\frac{1}{|\ln
(z/J_\infty)|}\;,
\end{equation}
with $b=3/2$. The correlation length $\xi$ is determined by the
equation for the density
\begin{equation}
\label{eq:phi} \phi = \sum_{n=0}^{\infty}{n
P(n)}/\sum_{n=0}^{\infty}{P(n)}\;.
\end{equation}
Since $b<2$ a finite correlation length $\xi$ may be found for any
density $\phi$. Thus, the distribution (\ref{eq:dist}) is valid
for any density and no phase separation takes place in this case.
On the other hand for $b>2$ the distribution (\ref{eq:dist}) can
support only sufficiently low densities, and therefore one expects
phase separation at high densities, with a macroscopic occupation
of one of the boxes, in analogy to Bose-Einstein condensation. To
this order, $\xi$ is determined by the density, and is independent
of $q$. Therefore the $q$ dependence of $\xi$ and the sharp
crossover observed in simulations must come from higher order
corrections to the current. As will be shown below, the main
effect of these corrections is to suppress the small domains. This
results in a large correlation length in the system.

\begin{figure}
\epsfysize 5 cm \epsfbox{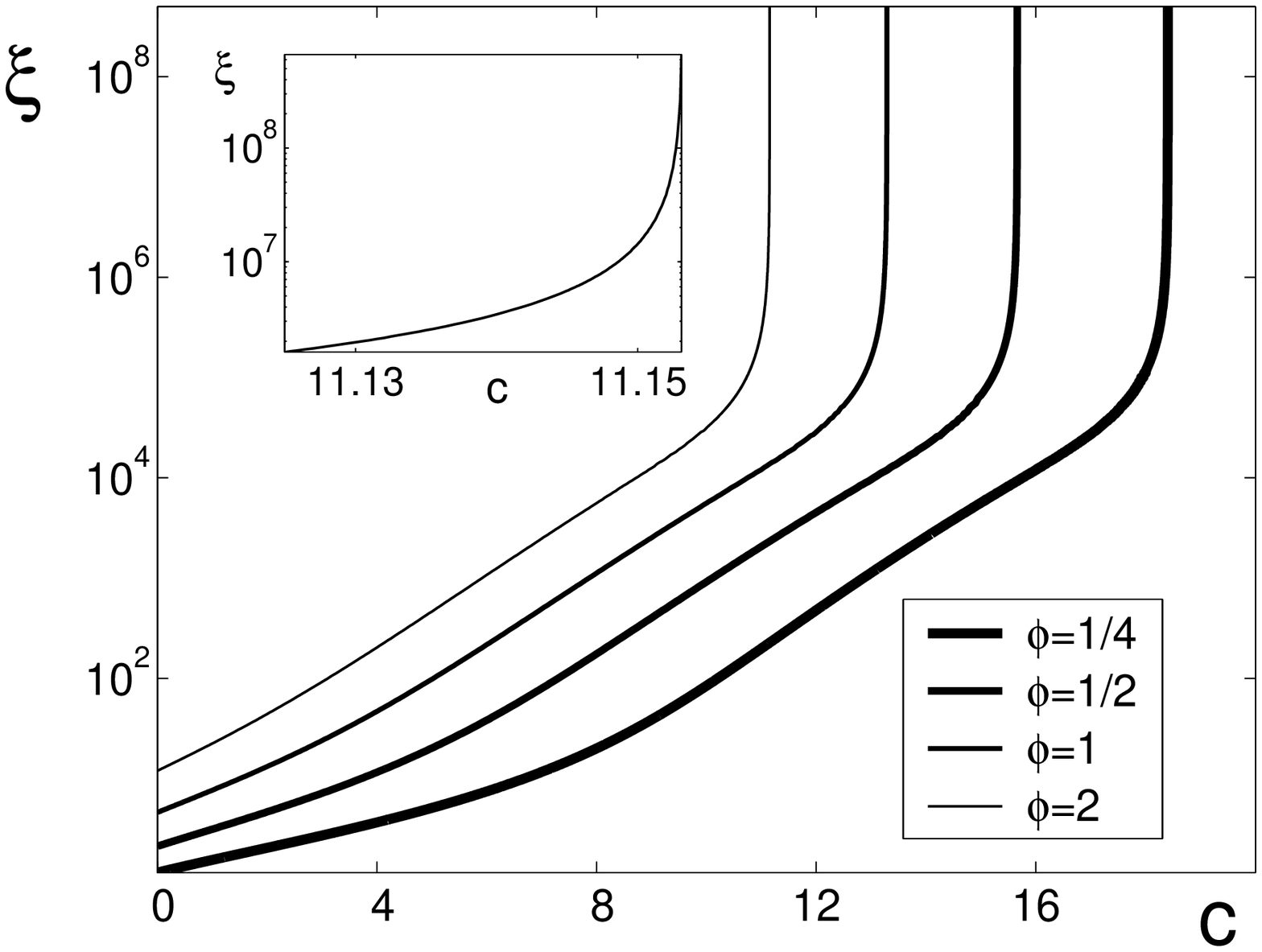} \epsfysize 5 cm
\epsfbox{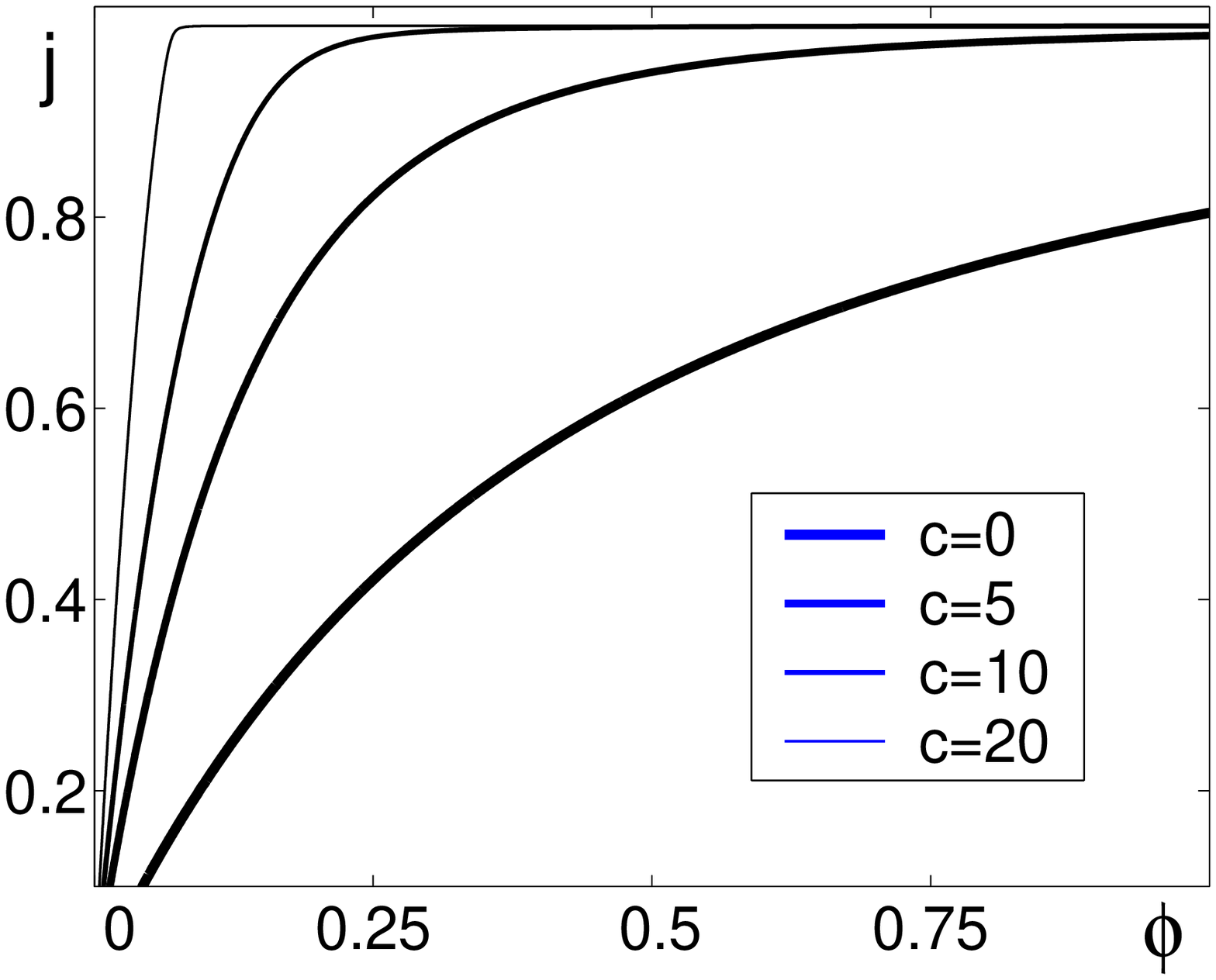} \caption{The correlation length $\xi$ as a
function of the coefficient $c$ for the ZRP with the transition
rates given by (\ref{eq:JTOY}). In the inset we zoom on the large
$\xi$ region for the case $\phi=2$, showing that in this region
$\xi$ grows faster than exponentially. Note that $\xi$ is plotted
on a logarithmic scale in both graphs.\label{fig:xiofc}}
\caption{The current $j$ as a function of $\phi$ for the same
model as in Fig. \ref{fig:xiofc}. For simplicity we set here
$J_\infty=1$. \label{fig:jofrho}}
\end{figure}

To demonstrate this point we first consider $J_n$ to order
$1/n^2$. We therefore study the ZRP with the currents $J_n$ given
by
\begin{equation}
\label{eq:JTOY} J_n =
J_\infty\left(1+\frac{b}{n}+\frac{c}{n^2}\right)\;.
\end{equation}
In making the correspondence to the AHR model we take $b=3/2$, and
note that to leading order in $(1-q)$ the current $J_n$ takes the
form~\cite{Sasamoto96}
\begin{equation}
J_n=\frac{1+(1-q)f(n,\alpha)}{n-1+2/ \alpha} \;,
\end{equation}
where $f(n,\alpha)$ is independent of $q$. Expanding this
expression in powers of $1/n$ we find that to leading order in
$(1-q)$, the coefficient $c$ diverges as $(1-q)^{-1}$. Thus, $c$
(and in fact all higher-order corrections) is $q$-dependent, and
becomes large near $q=1$. Using the form (\ref{eq:JTOY}) of the
current, the domains size distribution for large $n$
(\ref{eq:dist}) becomes
\begin{equation}
\label{eq:distTOY} P(n) \sim \frac{1}{n^b}\exp(-n/\xi-c/n)\;.
\end{equation}
The value of $\xi$ for a given density is determined by
(\ref{eq:phi}), and is therefore a function of $c$. In Fig.
\ref{fig:xiofc} we plot this function for several values of the
density $\phi$. For any density the correlation length $\xi$
exhibits two distinct regions in $c$ with a sharp crossover
between them but no singularity. For small $c$ the correlation
length seems to increase exponentially with $c$, while for large
$c$ the increase appears to be faster than exponential. This can
be seen in the inset of Fig. \ref{fig:xiofc}, where we zoom on a
very narrow range of values of $c$ at the crossover. The crossover
point increases with $\phi$. Since, as argued before, $c$ of the
AHR model diverges at $q=1$, we expect the crossover value to take
place for some $q<1$, as observed in direct numerical simulations
of the model and in accordance with the exact results.

We now note that a similar sharp crossover phenomenon also takes
place when one considers the behavior of the current as a function
of the density. Within the ZRP the quantity which corresponds to
the current of the AHR model is
\begin{equation}
j=\frac{\sum_n{n w_n P(n)}}{\sum_n{(n+1) P(n)}}\;,
\end{equation}
where $w_n$ is given by the current $J_n$ of (\ref{eq:JTOY}) to
order $1/n^2$ with $J_\infty=(1-q)/4$. It is straightforward to
show that in the grand-canonical ensemble $j=z$. Using this
relation, we plot $j$ as function of $\phi$ for fixed values of
$c$ in Fig. \ref{fig:jofrho}. Here again two distinct regions can
be observed for sufficiently large $c$. At low density $j$
exhibits a noticeable increase with $\phi$, while it slowly
approaches its maximal value $j=1$ at high densities. Between the
two regimes one finds a sharp crossover but no singularity. The
current density relation obtained here for the ZRP has the same
features as that of the AHR model (compare Fig. 2 of
\cite{Sasamoto} to Fig. \ref{fig:jofrho} here).


To get a better understanding of this behavior, we note that the
effect of large $c$ on the domain size distribution is to suppress
the weight of small domains. This results in a large correlation
length needed to sustain the density. In the AHR model, where the
correspondence to the ZRP is exact, this suppression is achieved
by large high-order corrections to the current $J_n$. From Fig.
\ref{fig:xiofc} it can be seen that for a given density the sharp
crossover occurs only for sufficiently large $c$. In this case it
is evident that the high-order corrections can take high enough
values as $q$ is increased, since these corrections diverge at
$q=1$. However, in other models this suppression may be induced by
other mechanisms, which do not require large high-order
corrections to the current of the form discussed above. To
demonstrate this point we consider a ZRP defined by the transition
rates
\begin{equation}
\label{eq:Wn0}
w_n=\left\{\begin{array}{lr} W & n<n_0\\
J_\infty(1+b/n) & n\geq n_0
\end{array}\right. \;,
\end{equation}
where $W$ and $n_0$ are free parameters, and $W$ directly controls
the weight of small domains. Sufficiently large $W$ destabilizes
small domains and suppresses them in the steady-state
distribution. This model with $b=0$ has been discussed previously
in the context of ZRP and demonstrated to exhibit a sharp
crossover by Evans~\cite{Evans} . A straightforward calculation
leads to the domain size distribution
\begin{equation}
\label{eq:distWn0}
P(n)\sim \left\{\begin{array}{lr} e^{-n/\xi_1} & n<n_0\\
 (n/n_0)^{-b}e^{-n_0/\xi_1}e^{-(n-n_0)/\xi_2} & n > n_0
 \end{array}\right.\;,
\end{equation}
with $\xi_1^{-1}=-\ln(z/W)$ and $\xi_2^{-1}=-\ln(z/J_\infty)$.
Note that the $n > n_0$ expression is valid only for large $n$. In
order to suppress small domains one needs $J_\infty<W$ which
yields $\xi_1<\xi_2$. Larger $n_0$ or $W$ result in a stronger
suppression of small domains, increasing the weight of the tail of
the distribution. We have calculated $\xi=|\ln(z)|^{-1}$ as a
function of $W$ for fixed density, and found that it exhibits
similar features to those of Fig. \ref{fig:xiofc}, with a sharp
crossover taking place at a value of $W$ which decreases with
$n_0$. For example, for $n_0=5$, the crossover takes place at $W
\simeq 8$. We thus conclude that the sharp crossover is a direct
result of suppression of small domains. The suppression of small
domains results in an increase of the tail of the domain size
distribution in order to keep the average density of particles.
This results in a larger correlation length. Of course, the
details of the suppression mechanism may differ from model to
model, and will affect the exact from of the distribution.

\begin{figure}
\epsfysize 5 cm \epsfbox{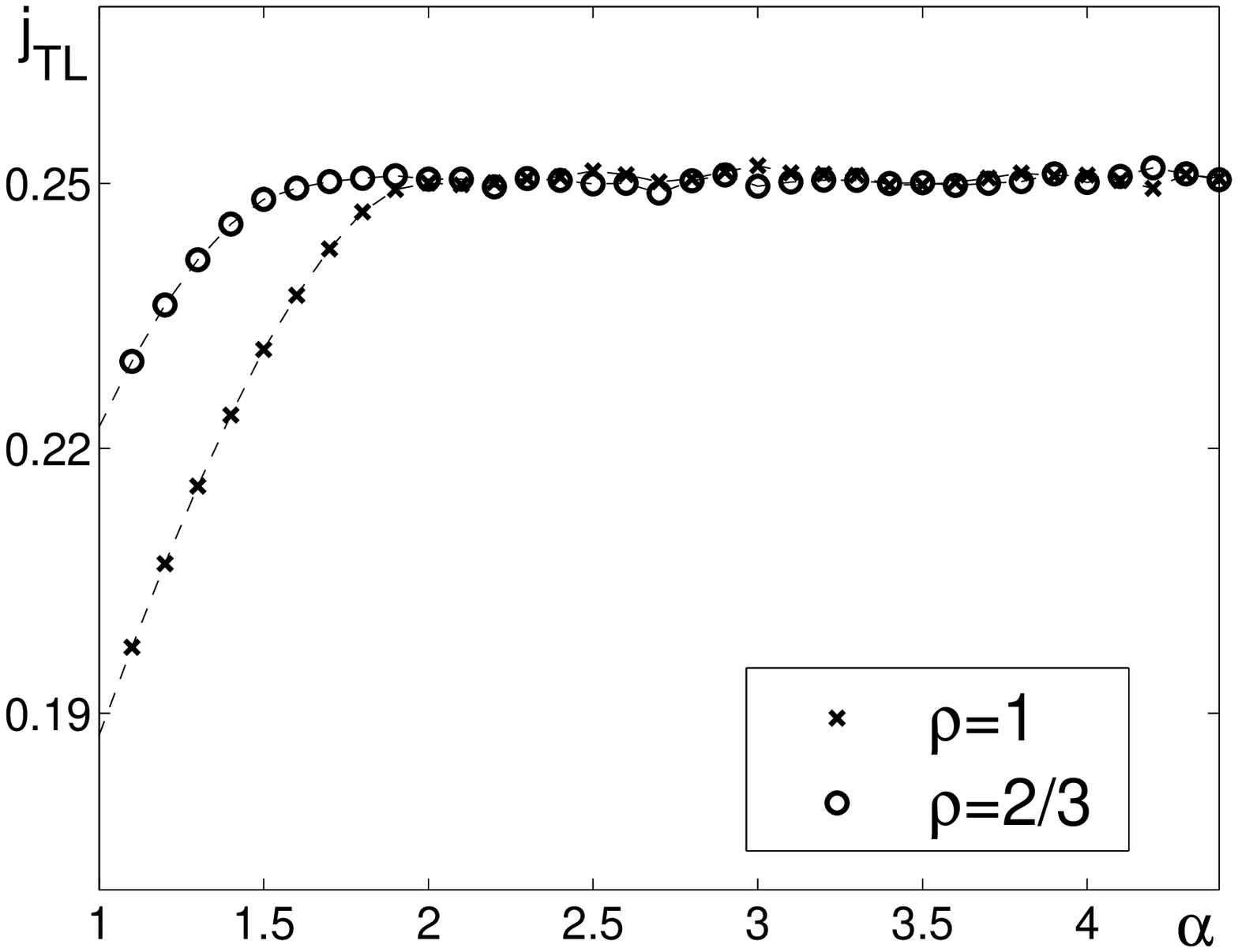} \epsfysize 5 cm
\epsfbox{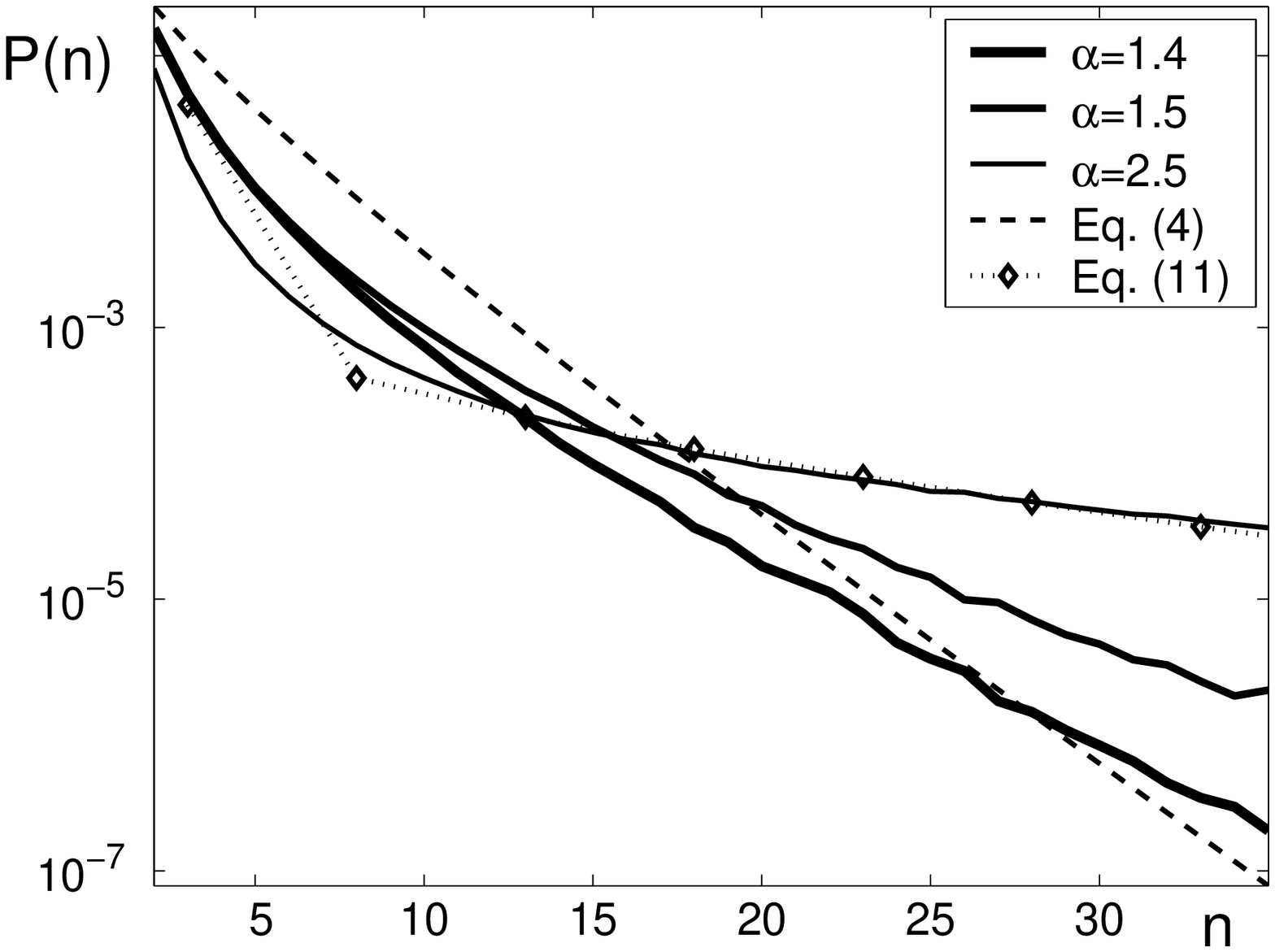} \caption{The current $j_{\mbox{\tiny TL}}$ of
positive particles in the two-lane model, as a function of
$\alpha$, for two densities. $j_{\mbox{\tiny TL}}$ was obtained
through direct numerical simulations of systems of size
5000.\label{fig:twolanecurr}}\caption{The domain size distribution
in the two-lane model for several values of the model parameter
$\alpha$. Here $\rho=1/2$ and $\gamma=1$. A domain of size $n$ is
defined as an uninterrupted sequence of length $n$ of particles
residing in both lanes. The normalization is such that domains of
size $0$ are not counted. Note that as $\alpha$ increases small
domains are more suppressed, and the tail of the distribution
becomes more pronounced. For comparison we plot the distribution
of the ZRP where small domains are not suppressed (dashed line),
and the one where small domains are suppressed as given by Eq.
(\ref{eq:distWn0}) (dotted line, here we arbitrarily take $W=4$
and $n_0=6$).
\label{fig:hist}}
\end{figure}

We now turn to consider the two-lane model introduced in
\cite{Korniss}. The model is defined on a $2\times L$ periodic
lattice, where each site is either empty, or occupied by a
positive or a negative particle. The dynamics within a lane is
governed by the rates (\ref{eq:rates}) with $q=0$. Particles are
also allowed to hop between lanes with the rates
\begin{equation}
\label{eq:twolanerates}
+\;0\mathop{\rightleftharpoons}\limits^{\gamma\alpha}_{\gamma\alpha}
0\;+ \;\;\;;\;\;\; -\,0
\mathop{\rightleftharpoons}\limits^{\gamma\alpha}_{\gamma\alpha}
0\,- \;\;\;;\;\;\; +\,-
\mathop{\rightleftharpoons}\limits^{\gamma}_{\gamma} -\,+ \;.
\end{equation}
Direct numerical simulations suggest that for large $\alpha$ the
model exhibits phase separation~\cite{Korniss}.

This model has recently been studied~\cite{Kafri} using its
conjectured correspondence to the ZRP, where the existence of a
phase transition is determined by the leading finite-size
correction to the current. It was suggested that the model does
not exhibit phase separation. In analogy with the analysis of the
AHR model, the current of a finite domain has been studied by
considering an open system of length $n$. This system is fully
occupied by positive and negative particles. In the interior the
dynamics follows the same rates of the model given by Eqs.
\ref{eq:rates} and \ref{eq:twolanerates}, although here there are
no vacancies. At the boundaries positive (negative) particles are
injected with rate $\alpha$ at the left (right). This
automatically implies that particles are removed at the other end
with the same rate $\alpha$. The dynamics within a domain is
therefore given by the particle exchange rates (\ref{eq:rates})
and (\ref{eq:twolanerates}) with $q=0$ in the interior, together
with the following rates at the boundaries of both lanes
\begin{equation}
\label{eq:opensystemrrates} -\mathop{\rightarrow}\limits^{\alpha}+
\quad\mbox{at site }1 \quad ; \quad
+\mathop{\rightarrow}\limits^{\alpha}- \quad \mbox{at site }n\;.
\end{equation}
This is a generalization of the totally asymmetric exclusion
process with open boundary conditions to two lanes. For this model
no analytical expression for the current $J_n$ of a domain of size
$n$ is available. Extensive numerical studies~\cite{Kafri} have
shown that the asymptotic form of the current is given by Eq.
\ref{eq:Jn} with $b \simeq 0.8$ for any $\gamma \neq 0$. Therefore
the model does not phase separate at any value of the parameters.
However, as in the AHR model, keeping the density fixed one
observes in numerical simulations of finite systems a sharp
crossover in the current as the parameter $\alpha$ is changed.
This can be seen, for example, in Fig. \ref{fig:twolanecurr},
where the current $j_{\mbox{\footnotesize TL}}$ of a two-lane
system is plotted as a function of $\alpha$. Here, too, the sharp
crossover seen in simulations may be misinterpreted as a phase
transition.

We now show that in the regime where phase separation seems to
take place, the statistical weight of small domains is suppressed,
in accordance with the mechanism suggested above. To this end we
have studied the domain size distribution in the two-lane model by
direct numerical simulations (see Fig. \ref{fig:hist}). It is
instructive to compare these results with the distribution
function (\ref{eq:dist}) with $b=0.8$, obtained from the ZRP where
small domains are not destabilized. One can see that small domains
in the two-lane system are significantly suppressed as compared
with this distribution. This suppression is easily understood
since small domains which form on one lane can readily dissolve
through the exchange to the other lane. As discussed above this
could lead to a sharp crossover in the correlation length. For
comparison we also plot the distribution function
(\ref{eq:distWn0}). Although one does not expect the details of
the distribution to agree it is evident that the general behavior
is captured.

In summary, a simple mechanism for the occurrence of a sharp
crossover to anomalously large correlation length exhibited in
some driven systems is suggested. The mechanism is examined in
detail for two models, and is shown to be consistent with
analytical results in one model, and with numerical results in the
other.

\ack We thank M. R. Evans and C. Godr\`eche for helpful
discussions. The support of the Israeli Science Foundation is
gratefully acknowledged.

\end{document}